\documentclass[aps,pra,twocolumn,superscriptaddress,showpacs,floatfix]{revtex4-1}
\usepackage{graphicx,amsmath,amssymb,amsfonts,multirow}

\usepackage[utf8]{inputenc}
\usepackage{color}

\newcommand{\ket}[1]{\mathinner{|{#1}\rangle}}
\newcommand{\bra}[1]{\mathinner{\langle{#1}|}}

\newcommand{\ketbra}[2]{\mathinner{|{#1}\rangle\langle{#2}|}}
\newcommand{\prj}[1]{\ketbra{#1}{#1}}
\newcommand{\mean}[1]{\mathord{\langle{#1}\rangle}}
\newcommand\diff[1]{\mathinner{\mathrm{d}#1}}
\renewcommand\Re{\mathop{\mathrm{Re}}}
\renewcommand\Im{\mathop{\mathrm{Im}}}

\begin{document}

\title{Spectral properties of single photons from quantum emitters} 

\author{Philipp M\"uller}
\affiliation{Experimentalphysik, Universit\"at des Saarlandes, 66123 Saarbr\"ucken, Germany}
\author{Tristan Tentrup}
\affiliation{Theoretische Physik, Universit\"at des Saarlandes, 66123 Saarbr\"ucken, Germany}
\author{Marc Bienert}
\affiliation{Theoretische Physik, Universit\"at des Saarlandes, 66123 Saarbr\"ucken, Germany}
\author{Giovanna Morigi}
\affiliation{Theoretische Physik, Universit\"at des Saarlandes, 66123 Saarbr\"ucken, Germany}
\author{J\"urgen Eschner}
\thanks{\href{mailto:juergen.eschner@physik.uni-saarland.de}{juergen.eschner@physik.uni-saarland.de}}
\affiliation{Experimentalphysik, Universit\"at des Saarlandes, 66123 Saarbr\"ucken, Germany}

\date{\today}

\begin{abstract}
Quantum networks require flying qubits that transfer information between the nodes. This may be implemented by means of single atoms (the nodes) that emit and absorb single photons (the flying qubits) and requires full control of photon absorption and emission by the individual emitters. In this paper, we theoretically characterize the wave packet of a photon emitted by a single atom undergoing a spontaneous Raman transition in a three-level scheme. We investigate several excitation schemes that are experimentally relevant and discuss control parameters that allow one to tailor the spectrum of the emitted photon wave packet. 
\end{abstract}

\maketitle


\section{\label{Introduction}Introduction}

Absorption and emission of photons by atoms are the fundamental processes of light-matter interaction~\cite{Cohenbook}. At the same time they form the basic building blocks of quantum networks~\cite{Cirac1997, Kimble2008, Ritter2012, Duan2010} that consist of atomic nodes and photons carrying information between them. In several protocols, the information is transferred directly between the atoms by controlled photon emission and absorption \cite{Cirac1997}. In other schemes information processing is achieved by means of projective measurements via photo-detection \cite{Cabrillo1999, Plenio1999, Simon2003, Feng2003}. Either way, a fundamental requirement is the control of the spectral and temporal properties of single photons that are released from a single emitter through controlled excitation~\cite{Keller2004, Darquie2005, Almendros2009, Kurz2013, Nisbet2013, Schug2014}. These properties determine the absorption probability by a single atom \cite{Stobinska2009} as well as the interference contrast of photon--photon (or Hong--Ou--Mandel) interference \cite{Beugnon2006, Maunz2007, Gerber2009}, which is utilized to entangle remote atoms \cite{Moehring2007, Hofmann2012, Bernien2013}.

\begin{figure}[htb]
	\centering
	\includegraphics{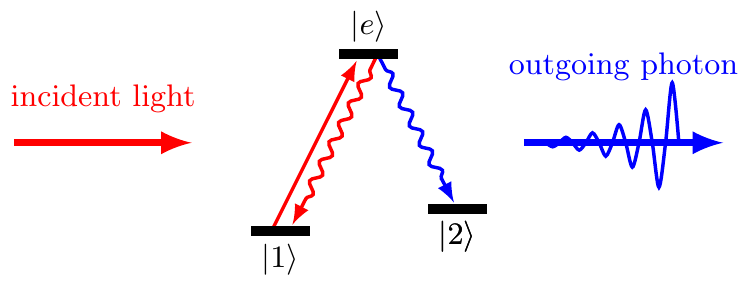}
	\caption{\label{fig:model} An atom with a $\Lambda$-shaped level configuration is prepared in state~$\ket{1}$ and excited to state~$\ket{e}$ by incident light. A single photon may be emitted along the transition $\ket{e}\to\ket{2}$. We determine the spectral properties of the emitted photon as a function  of the atomic parameters and of the properties of the incident light.}
\end{figure}

Raman transitions, such as the one sketched in fig.\,\ref{fig:model} are particularly relevant for controlled single-photon creation. The three-level design is convenient in order to separate excitation and emission, and to terminate the dynamics after the creation of one desired single photon. Moreover, the created single photon may be entangled with the emitting atom~\cite{Blinov2004}. A generic situation is that incident laser light releases the single photon. Alternatively, it may happen by single photons that themselves are created from quantum emitters~\cite{Rezus2012, Schug2013, Aharonovich2016, Delteil2017} or other single-photon sources \cite{Fasel2004, Fedrizzi2007, Bao2008, Haase2008, Scholz2009}. This is an interesting case of atom-photon interface, as the emitted photon allows one to herald the absorption process \cite{Piro2011, Schug2013, Schug2014, Lenhard2015, Brito2016, Kurz2016}.

In this paper we study theoretically the wave-packet properties of single photons that are generated in a spontaneous Raman scattering process in an atomic three-level system. We consider excitation of the atom by single photons of various spectro-temporal properties, and by laser light. We pay particular attention to the coherence of the photon, i.\,e.\ its time--bandwidth product. Moreover, we include the effect of the branching ratio of the upper atomic level. With respect to previous work as in refs.\,\cite{Knight1980, Zhu1995, Zhu1995a}, we determine the single-photon spectrum produced by a generic excitation and consider the details of the atomic level configuration, such as a possible decay back to the initial state of the Raman transition. Although the spectral shape for the most efficient absorption of a single photon has been studied before \cite{Wang2011, Leong2016}, those studies were restricted to two-level systems. Our study allows us to determine the properties of the photon emitted in a Raman transition as a function of the excitation parameters and the atomic properties, and thus to identify the perspectives for controlling its shape in cases of experimental relevance.

This paper is organized as follows. In Sec.\,\ref{sec:model} we introduce the model, which we apply in Sec.\,\ref{sec:spectrum} in order to determine the spectral properties of the emitted photon for various cases of an exciting field driving the Raman transition. In Sec.\,\ref{sec:beats} we apply our results to the description of quantum beats, recently reported in~\cite{Schug2014}. The conclusions are drawn in Sec.\,\ref{sec:conclusion}, and the appendices provide further theoretical details complementing the calculations in Sec.\,\ref{sec:spectrum}.

\section{Model} 
\label{sec:model}

In this section we introduce the theoretical model and the basic equations from which we determine the spectrum of the emitted photon as a function of the properties of the incident light. 

\subsection{Hamiltonian}

We consider three electronic levels~$\ket{1}$, $\ket{2}$, and $\ket{e}$ of a single atom forming a $\Lambda$-configuration, as illustrated in fig.\,\ref{fig:model}. Each stable state~$\ket{j}$ ($j=1,2$) is connected to the excited state~$\ket{e}$ by an optical dipole transition of frequency~$\omega_{ej}$. The two corresponding radiation fields are distinguishable, for example by their polarizations and/or their frequencies. We will describe a scattering process where incident light on the transition $\ket{1}\leftrightarrow\ket{e}$ excites the atom and spontaneous decay on the transition $\ket{e}\leftrightarrow\ket{2}$ creates a single-photon wave packet. 

We start with the Hamiltonian~$H$ of atom and light fields, treating the dipolar emission pattern as a single spatial mode.  We use the bosonic operators $b_j(\omega)$ and $b_j^+(\omega)$ for the fields coupling to the transitions $\ket{j}\leftrightarrow\ket{e}$ to denote the annihilation and creation of a photon with frequency~$\omega$ and wave number $k=\omega/c$, whereby $[b_{j}(\omega),b_{j'}^+(\omega')] = \delta_{jj'}\delta(\omega-\omega')$.  We decompose~$H$ into the free part~$H_0$ and the atom--field interaction part~$W$,
\begin{equation}\label{Hamiltonian}
	H = H_0+W.
\end{equation}
In detail,
\begin{equation}\label{eq:H0}
	 H_0 =- \sum_{j=1}^2\hbar \omega_{ej}\prj{j} + \hbar\int\limits_0^\infty \diff\omega \omega b_{j}^+(\omega) b_{j}(\omega)\,,
\end{equation}
whereby the first and second terms represent the energy of the atomic states and of the radiation field, respectively, and the energy of the excited state~$\ket{e}$ has been set to zero.
The atom--photon interaction $W=W_1+W_2$ is composed of the terms~$W_1$ and $W_2$ that describe the coupling to the fields $b_1(\omega)$ and $b_2(\omega)$, respectively, in electric-dipole approximation,
\begin{align}\label{eq:W}
	W_j = \hbar \int\limits_0^\infty \diff\omega \sqrt{\frac{\Gamma_j}{2\pi}} \ketbra{e}{j}b_{j}(\omega)+\text{H.\,c.}
\end{align}
Treating the dipolar wave pattern as a single mode allows us to express the atom-photon coupling constant in Eq.\,(\ref{eq:W}) directly through the Einstein $A$-coefficients, $\Gamma_j$, of the two transitions $\ket{e}\to\ket{j}$ in the Weisskopf--Wigner approximation.  Here, $\Gamma = \Gamma_1 + \Gamma_2$ is the spontaneous decay rate of the excited state.

\subsection{Scattering amplitude}

The spectral properties of the emitted photon are calculated by determining the transition amplitude~$U_{fi}(\omega_2,t)$ from the initial state $\ket{i}=\ket{\Psi(0)}=\ket{1;\varphi}$, i.\,e.\ atom in state~$\ket{1}$ and field in state~$\ket{\varphi}$, into the target state 
\begin{equation}\label{eq:f}
	\ket{f} = b_2^+(\omega_2) \ket{2;\text{vac}},
\end{equation}
that corresponds to the atom in state~$\ket{2}$ and a photon in the mode at frequency~$\omega_2$. (We will also consider other target states when discussing the effect of the finite branching ratio on the spectral-temporal properties of the emitted photon.) Under coherent time evolution with Hamiltonian~$H$, the transition amplitude (for $t>0$) is given by
\begin{align}\label{eq:Ufi}
	U_{fi}(\omega_2,t) &= \bra{f}U(t)\ket{\Psi(0)} \nonumber\\
	&= \frac{1}{2\pi i} \int\limits_{C_+} \diff z e^{-\frac{izt}{\hbar}}\bra{f}G(z)\ket{\Psi(0)},
\end{align}
whereby
\begin{equation}\label{eq:Gofz}
	G(z)= \frac{1}{z-H}
\end{equation}
is the analytic extension of the propagator to the complex plane, and $C_+$ is the contour for $z=E+i\eta$ with energy~$E$ varying from $+\infty$ to $-\infty$, and $\eta\to 0^+$. We calculate the matrix elements by means of the Wigner--Weisskopf approximation and determine the probability density that a single photon at time~$t$ and frequency~$\omega_2$ is generated:
\begin{align}\label{eq:Pomegat}
	{\mathcal P}(\omega_2,t) = |U_{fi}(\omega_2,t)|^2\,.
\end{align}
From this quantity we extract the area-normalized power spectrum of the emitted photon,
\begin{align}\label{eq:Somegat}
	{\mathcal S}(\omega_2,t) &= \frac{1}{\mathcal N(t)}{\mathcal P}(\omega_2,t)\,,
\end{align}
and the probability that a photon is emitted along the transition $\ket{e}\to\ket{2}$,
\begin{equation}\label{eq:calN}
	\mathcal N(t)=\int\limits_0^{\infty} {\mathcal P}(\omega_2,t) \diff\omega_2\,,
\end{equation}
We interpret $\mathcal N(t)$ as the accumulated success probability for the creation of a photon of any frequency during the interaction time~$t$. 

The spectral properties depend on the state of the input field $\ket{\varphi}$. For a single photon this takes the generic form
\begin{equation}\label{eq:i}
	\ket{\varphi} = \int\limits_0^\infty \diff\omega \psi(\omega)b_1^+(\omega) \ket{\text{vac}}\,,
\end{equation}
where $\ket{\text{vac}}$ is the vacuum state of the electromagnetic field and $\psi(\omega)$ is the probability amplitude distribution in frequency, with $\int_0^\infty \diff\omega |\psi(\omega)|^2=1$. For a single-mode c.\,w.\ laser, $\ket{\varphi}$ is a coherent state; we will discuss this case in Sec.\,\ref{Laser}. In the single-photon case, we write the final expression for ${\mathcal P}(\omega_2,t)$ as
\begin{align}\label{eq:P}
	{\mathcal P}(\omega_2,t) = \left|\int\limits_{0}^\infty \diff\omega \psi(\omega) u(t, \omega, \omega_2)\right|^2\,,
\end{align}
where $u(t,\omega,\omega_2) = \langle 2,\text{vac}|b_2(\omega_2)U(t)b_1^+(\omega)|1,\text{vac}\rangle$ is calculated according to Eq.\,\eqref{eq:Ufi} and \eqref{eq:Gofz}. The quantity $u(t,\omega,\omega_2)$ is thus the probability amplitude that a monochromatic photon of frequency~$\omega$ is absorbed from field~1 and a photon of frequency~$\omega_2$ is emitted into field~2. Equation~\eqref{eq:P} shows that this matrix element is the building block needed for calculating the spectrum~${\mathcal S}(\omega_2,t)$, Eq.\,\eqref{eq:Somegat}. Using resolvent theory and the Wigner--Weisskopf approximation~\cite{Cohenbook}, we find
\begin{multline}\label{eq:ufi}\textstyle
	u(t, \Delta_1, \Delta_2) = \frac{\sqrt{\Gamma_1\Gamma_2}}{2\pi} \left[ \frac{\displaystyle e^{-i\Delta_2 t}}{\left( \Delta_2 +i\frac{\Gamma}{2} \right)\left( \Delta_2-\Delta_1 \right)} \right. \\
	\textstyle
	+ \left.\frac{\displaystyle e^{-\frac{\Gamma}{2}t}}{\left( \Delta_1+i\frac{\Gamma}{2} \right)  \left( \Delta_2 +i\frac{\Gamma}{2} \right)} +\frac{\displaystyle e^{-i\Delta_1 t}}{\left( \Delta_1 +i\frac{\Gamma}{2} \right)\left( \Delta_1 -\Delta_2\right)} \right]\,,
\end{multline}
where $\Delta_1 = \omega_1-\omega_{e1}$ and $\Delta_2 = \omega_2-\omega_{e2}$ are the photon frequencies shifted by the value of the corresponding transition frequency. The shift of the excited state due to virtual photon processes is absorbed in the definition of its energy.

\section{Spectrum of the emitted photon}
\label{sec:spectrum}

In this section we determine and discuss the spectral form of the emitted photon by spontaneous emission on the transition $\ket{e}\to\ket{2}$, triggered by (i)~a single incident photon that excites the transition $\ket{1}\leftrightarrow\ket{e}$ or (ii)~by a laser continuously driving the transition $\ket{1}\leftrightarrow\ket{e}$. 

We note that all the expressions of interest, and in particular the spectrum of the emitted photon, depend on the observation time interval~$t$. Although the full time dependence may be derived with our method, we will restrict ourselves to the limit $t\to\infty$. This means we look at the spectrum of the emitted photon in the asymptote of the emission process. 

\subsection{Excitation by a single photon}

\subsubsection{Rectangular wave packet}

A relevant realistic photonic state is a rectangular pulse with a monochromatic carrier. For a pulse duration~$T$ the wave packet is described by
\begin{equation*}
	\tilde\psi(t) = \frac{e^{-i\omega_1 t}}{\sqrt T} \Theta(t-\tau) \Theta(\tau+T-t),
\end{equation*}
where $\tau$ denotes the initial temporal distance from the atom, such that the front of the wave packet is at distance~$x=c\tau$ from the scatterer's position. The amplitude spectrum reads
\begin{equation}\label{sincpacket}
	\psi(\omega) = e^{i(\omega-\omega_1)(\tau+\frac{T}{2})} \mathinner{\sqrt{\frac{2\pi}{T}}} \delta^{(\frac{T}{2})} (\omega-\omega_1)\,,
\end{equation}
where we used the diffraction function,
\begin{equation}\label{diffractionfunction}
	\delta^{(t)}(x) = \frac{\sin xt}{\pi x},
\end{equation}
which converges towards the Dirac delta distribution~$\delta(x)$ for $t\to\infty$.

The spectral shape of the emitted photon follows by inserting Eq.\,\eqref{sincpacket} into Eq.\,\eqref{eq:Somegat} and subsequently taking the limit $t\to\infty$. This yields (for $\tau, T>0$)
\begin{equation*}
	\mathcal{S}(\Delta_2) = \frac{2\pi \Gamma_1 \Gamma_2}{\mathcal N \Gamma} L(\Delta_2) \frac{2\pi}{T} \left( \delta^{(\frac{T}{2})} (\Delta_2-\Delta_1) \right)^2,
\end{equation*}
which is the product of the sinc-shaped spectral amplitude of the incident photon, with center frequency $\Delta_2 = \Delta_1$ and spectral linewidth $\Delta\omega_1 = 2\sqrt3/T$, and the atomic Lorentzian~$L(\Delta_2)$ of linewidth~$\Gamma$,
\begin{equation}
	L(\Delta) = \frac{\frac{\Gamma}{2\pi}}{\Delta^2+\left( \frac{\Gamma}{2} \right)^2}.
\end{equation}

\begin{figure}[htb]
	\centering
	\includegraphics{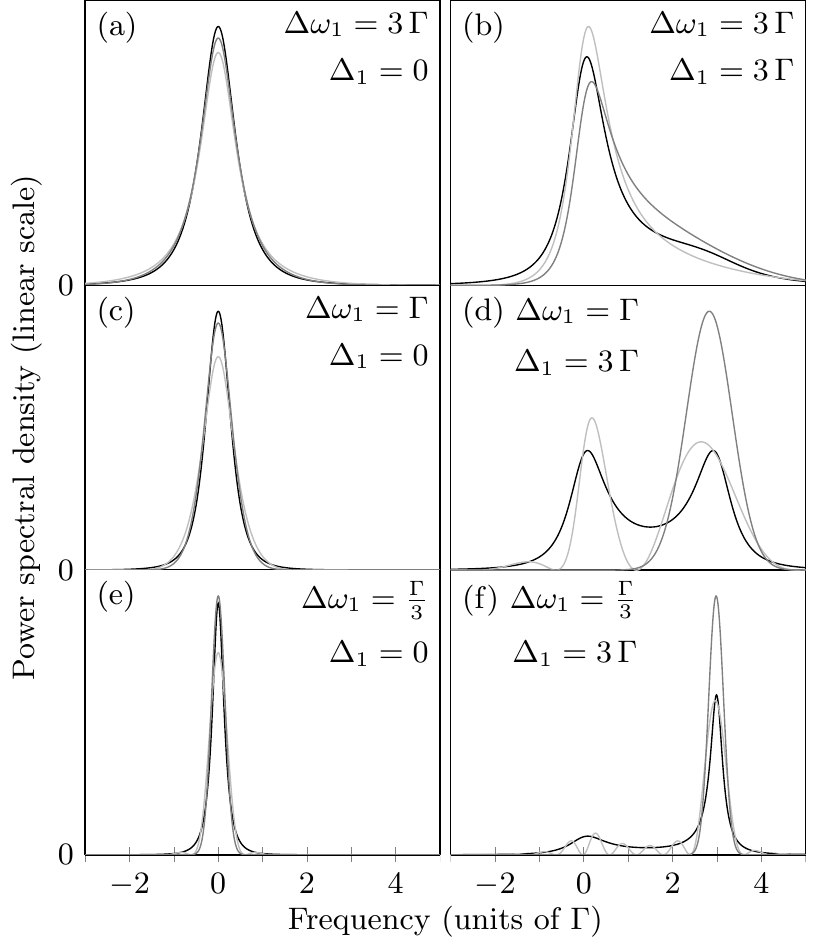}
	\caption{\label{fig:spectrumgauss}Area-normalized power spectral density~$\mathcal S(\Delta_2)$ of the emitted photon after excitation by a single photon of sinc (light gray), Gaussian (gray), Lorentzian (black) spectrum. The spectra are given for three different linewidths~$\Delta\omega_1$ and for resonant excitation in (a), (c), and (e), and an off-resonant excitation in (b), (d), and (f). 
	}
\end{figure}

\begin{figure}[htb]
	\centering
	\includegraphics{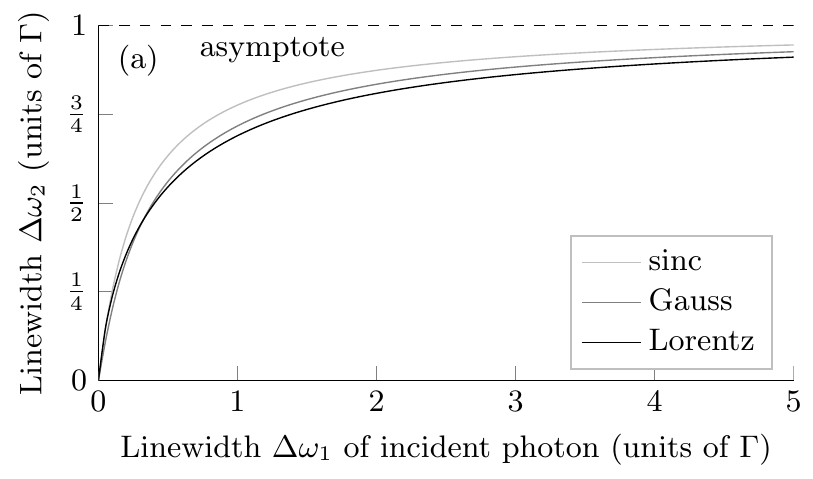}
	\par 
	\includegraphics{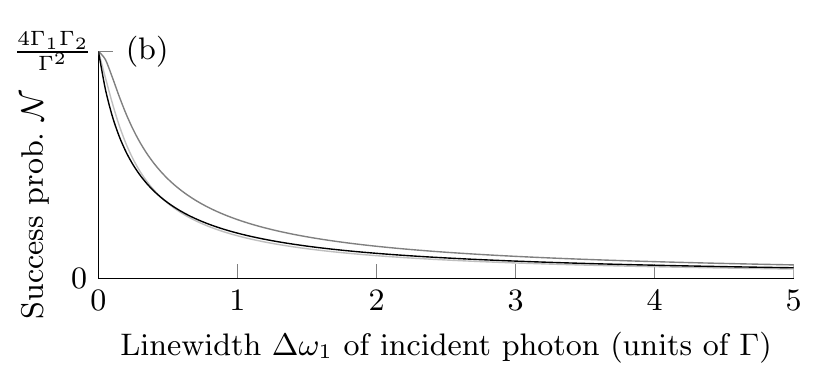}	\caption{\label{fig:Gaussvarianzbreite}(a)~Effective linewidth~$\Delta\omega_2$ and (b)~success probability~$\mathcal N$ of the emitted single photon versus the linewidth of the incident photon. Shown are the three cases of excitation with a single photon of sinc- (light gray), Gaussian- (gray), or Lorentzian-shaped (black) spectra at resonance ($\Delta_1=0$) and interaction time~$t \to \infty$. For all cases, the asymptote is $\Delta\omega_2 = \Gamma$ corresponding to the bare atomic Lorentzian.}
\end{figure}

The power spectrum of the emitted photon is plotted in fig.\,\ref{fig:spectrumgauss} (gray curves) for different values of $\Delta_1$ and $\Delta\omega_1$. To quantify its linewidth, we use an effective value,
\begin{equation}\label{sueszmann}
	\Delta\omega_2 = \frac{\delta_S}{\pi} = \frac{\left( \int \diff{\omega_2} \mathcal S(\omega_2) \right)^2}{\pi \int \diff{\omega_2} \mathcal S^2(\omega_2)}
\end{equation}
expressed through the Süßmann measure~$\delta_S$~\cite{Schleichbook}. Its dependence on the incident linewidth is shown in fig.\,\ref{fig:Gaussvarianzbreite}(a) (as a gray curve) for resonant excitation, $\Delta_1 = 0$. One sees that $\Delta\omega_2$ depends on the total atomic linewidth~$\Gamma$, but not on the branching fractions, $\Gamma_1$ and $\Gamma_2$, individually (at least for $t \to \infty$). Figure~\ref{fig:Gaussvarianzbreite}(b) shows the success probability for creating a single photon, as a function of the linewidth of the incident photon. 

\subsubsection{Gaussian wave packet}\label{sec:phwavepacket}

As another example for a photonic wave packet we consider the normalized Gaussian spectrum
\begin{equation}\label{Gausspacket}
 \psi(\omega)= \sqrt[4]{\frac{2}{\pi \Delta \omega^2}}\exp\left[{-\frac{\left(\omega-\omega_1\right)^2}{\Delta\omega_1^2}}\right]e^{i\left(\omega-\omega_1 \right) \tau}
\end{equation}
with central frequency~$\omega_1$, linewidth~$\Delta\omega_1$, and initial temporal distance~$\tau$ from the atom. Note that this Gaussian does not represent an incoherently broadened photon, but rather a pure (i.\,e.\ Fourier-limited) state~\footnote{We identify a pure photonic quantum state with a Fourier-limited wave packet, i.\,e.\ a wave packet whose temporal envelope and spectrum are Fourier transforms of each other. This is obviously the case if the quantum state is represented by a wave function~$\psi$ as in Eq.\,\eqref{Gausspacket} but no longer if temporal or spectral broadening processes exist that require incoherent summation over these processes weighted with their probabilities as in Eq.\,\eqref{eq:spectrum}.}. The spectral shape of the emitted photon follows again by multiplying Eq.\,\eqref{eq:ufi} with the wave packet, Eq.\,\eqref{Gausspacket}. Hence, in the limit $t\to\infty$ and for long distances~$\tau \gg \frac{1}{\Delta\omega_1}$, the power spectrum is the product of the atomic Lorentzian and a Gaussian function of width~$\Delta\omega_1$ centered at $\Delta_2 = \Delta_1$,
\begin{equation*}
	\mathcal{S}(\Delta_2) = \frac{2\pi \Gamma_1 \Gamma_2}{\mathcal{N}\Gamma} L(\Delta_2) \sqrt{\frac{2}{\pi\Delta\omega_1^2}} \exp\left[-\frac{2(\Delta_2-\Delta_1)^2}{\Delta\omega_1^2}\right].
\end{equation*}
The spectral shape, the effective linewidth, and the success probability of the emitted photon are displayed (as light-gray curves) in figs.\,\ref{fig:spectrumgauss} and \ref{fig:Gaussvarianzbreite}, respectively.

\subsubsection{Lorentzian wave packet}\label{sec:lwavepacket}

Another relevant example is a Lorentzian wave packet of linewidth~$\Delta\omega_1$, for example, when the photon is released from a cavity or from a single atom \cite{Almendros2009, Kurz2013}. Then, 
\begin{equation*}
	\psi(\omega)= \sqrt{\frac{\Delta\omega_1}{2\pi}}\frac{1}{(\omega-\omega_1)+i\frac{\Delta\omega_1}{2}}e^{i(\omega-\omega_1)\tau},
\end{equation*}
and the spectrum that follows from Eq.\,\eqref{eq:Somegat} reads (again for $\tau>0$ and $t\to\infty$) 
\begin{equation*} 
	\mathcal{S}(\Delta_2) = \frac{2\pi\Gamma_1\Gamma_2}{\mathcal N \Gamma} L(\Delta_2) \frac{\frac{\Delta\omega_1}{2\pi}}{(\Delta_2-\Delta_1)^2+(\frac{\Delta\omega_1}{2})^2}.
\end{equation*}
It is the product of two Lorentzians of widths~$\Gamma$ and $\Delta\omega_1$, centered at $\Delta_2=0$ and $\Delta_2=\Delta_1$, respectively. Examples are displayed in fig.\,\ref{fig:spectrumgauss} (as black curves).

In this case, the success probability according to Eq.\,\eqref{eq:calN},
\begin{equation*}
	\mathcal N = \frac{\Gamma_1\Gamma_2}{\Gamma} \frac{\Gamma+\Delta\omega_1}{\Delta_1^2+\left( \frac{\Gamma+\Delta\omega_1}{2} \right)^2}, 
\end{equation*}
is maximal for resonant excitation ($\Delta_1=0$) by a narrow-band photon ($\Delta\omega_1 \to 0$), and it reaches unity (in our one-dimensional model) for  the case of equal branching fractions, $\Gamma_1=\Gamma_2=\frac{\Gamma}{2}$.
The spectral width and the success probability behave similarly as in the case of Gaussian excitation, see fig.\,\ref{fig:Gaussvarianzbreite}.

We note that a time-reversed Lorentzian photon, i.\,e.\ one with an exponentially rising temporal envelope~\cite{Srivathsan2014}, has the same power spectrum and thus produces the same emission spectrum~$\mathcal S(\Delta_2)$ as calculated above. In the time domain, however, where the wave packet of the emitted photon is the convolution of the incident wave packet and the exponential atomic response, the two cases will lead to different results~\cite{Wang2011, Leong2016}.

\subsection{\label{Laser}Laser excitation}

Instead of a single photon which excites the transition $\ket{1}\leftrightarrow\ket{e}$, now we consider a laser with frequency~$\omega_1$ driving this transition continuously. The laser drive is assumed to be monochromatic and represented by a coherent state of the corresponding mode of the electromagnetic field. In an equivalent reference frame, $\ket{\varphi}=\ket{\rm vac}$, and the Hamiltonian reads \cite{Cohenbook}
\begin{equation}\label{eq:HL}
	H' = H_0' + W_1 + W_2 + V,
\end{equation}
where the free Hamiltonian~$H_0'$ corresponds to Eq.\,\eqref{eq:H0} with the eigenfrequency of the initial atomic state shifted by the laser frequency, $-\omega_{e1} \to -\omega_{e1}+\omega_1 = \Delta_1$, and the interaction of the laser and the atom is described by 
\begin{equation}\label{eq:V}
	V = \frac{\hbar\Omega}{2} \left( \ketbra{1}{e}+\ketbra{e}{1} \right)
\end{equation}
with the on-resonance Rabi frequency~$\Omega$. 

\begin{figure}[htb]
	\centering
	\includegraphics{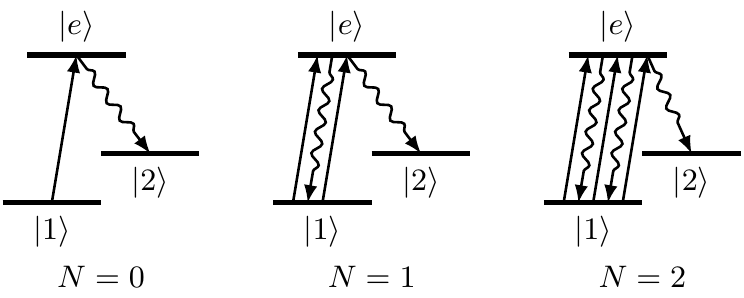}
	\caption{\label{threecases} Three cases of laser-induced generation of a single photon on the transition $\ket{e}\to\ket{2}$ with $N=0$,~1,~or~2 additional spontaneously emitted photons on the transition $\ket{1}\leftrightarrow\ket{e}$.}
\end{figure}

In contrast to the case of single-photon excitation, the final state now includes the possibility that multiple photons have been emitted spontaneously along the transition $\ket{e}\to\ket{1}$ before the photon emitted on $\ket{e}\to\ket{2}$ terminates the process. This is illustrated in fig.\,\ref{threecases}. For each case of $N$~such additional photons, the final state reads
\begin{equation*}
\ket{f_N} = b^+_2(\omega_2) b^+_1(\omega_1^{(N)})\ldots b^+_1(\omega_1^{(1)})\ket{2; {\rm vac}},
\end{equation*}
and the spectrum of the corresponding outgoing photon is given by
\begin{equation*}
\mathcal S_N(\omega_2, t) = \frac{1}{\mathcal N_N} \left|U_{f_N i}(t) \right|^2,
\end{equation*}
with the transition amplitude $U_{f_N i}(t) = \bra{f_N}U(t)\ket{\Psi(0)}$ calculated using the resolvent of $H'$, thus using Eq.\,\eqref{eq:HL} in Eq.\,\eqref{eq:Ufi}.

\begin{figure}[bth]
	\centering
	\includegraphics{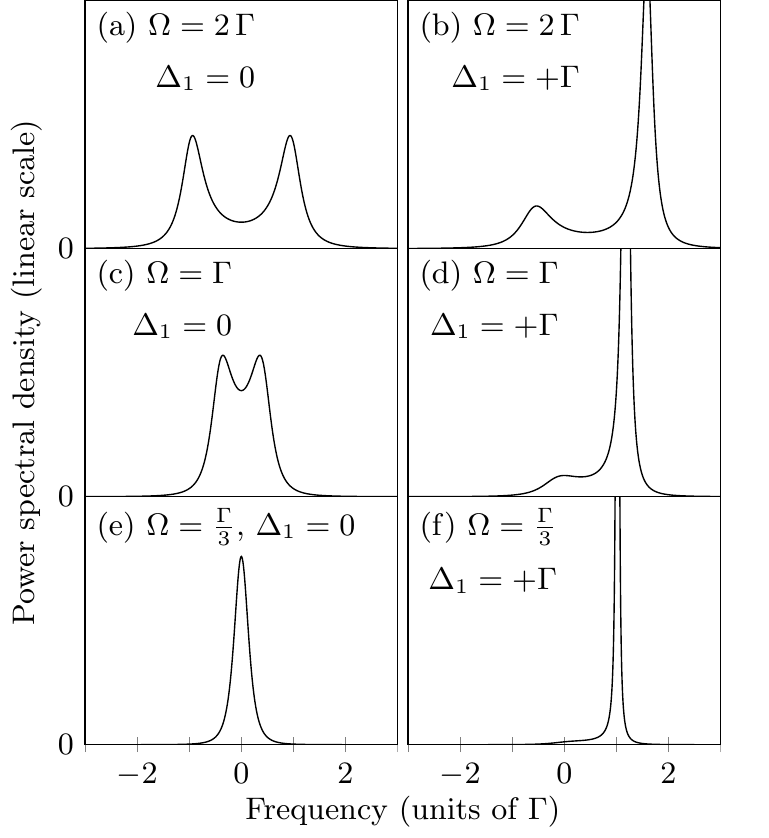}
	\caption{\label{fig:spectrumlaser}Area-normalized power spectral density~$\mathcal S_0(\Delta_2)$, Eq.\,\eqref{spectrum:0}, of the emitted single photon after excitation by a laser for various values of Rabi frequency~$\Omega$ and laser detuning~$\Delta_1$.}
\end{figure}

The full spectrum of the emitted photon is given by the incoherent sum over all possible cases of $N$ additional photon emissions, including the integration over their frequencies (this reflects, in practical terms, the assumption that all information about the additional photons is discarded) and weighted by the corresponding probability~${\mathcal N_N}$,
\begin{equation}\label{eq:spectrum}
\mathcal S (\omega_2, t) = \sum_{N=0}^\infty \mathcal N_N \mathcal S_N(\omega_2, t),
\end{equation}
where 
\begin{equation}\label{eq:norm}
	\sum_N {\mathcal N_N} = 1. 
\end{equation}
The first three processes of the sum, for $N=0$,~1,~2, are shown in fig.\,\ref{threecases}. For all three cases we obtain, as detailed in appendix~\ref{app_laser},
\begin{multline}\label{spectrum:0}
	\lim\limits_{t\to\infty}S_N(\omega_2,t) = \mathcal{S}_0(\Delta_2)\\
	= \frac{\frac{|\Omega|^2}{4} \frac{\Gamma}{2\pi} }{\left| \left( \Delta_2-\Delta_1-\Delta_S+i\frac{\kappa}{2} \right)\left( \Delta_2+\Delta_S+i\frac{\Gamma-\kappa}{2} \right) \right|^2}, 
\end{multline}
which is a product of two Lorentzian functions centered at $\Delta_2 = \Delta_1+\Delta_S$ and $\Delta_2=-\Delta_S$, and with widths
\begin{equation*}
	\kappa = \Gamma\frac{\Delta_S}{\Delta_1+2\Delta_S}
\end{equation*}
and $\Gamma-\nobreak\kappa$, respectively. Therein, $\Delta_S$ is the AC Stark shift due to the laser field,
\begin{equation*}
	\textstyle\Delta_S = -\frac{\Delta_1}{2}+\frac{\mathop{\text{sgn}}\Delta_1}{2\sqrt{2}} \sqrt{ \tilde\Omega^2-\frac{\Gamma^2}{4}+\sqrt{\left( \tilde\Omega^2-\frac{\Gamma^2}{4} \right)^2+\Delta_1^2 \Gamma^2}},
\end{equation*}
with the effective Rabi frequency~$\tilde\Omega = \sqrt{|\Omega|^2+\Delta_1^2}$. The shape of the spectrum according to Eq.\,\eqref{spectrum:0} is plotted in fig.\,\ref{fig:spectrumlaser}. It exhibits the well-known feature of Autler--Townes splitting at sufficiently high Rabi frequency, $\Omega > \frac\Gamma2$.

The corresponding success probabilities are $\mathcal N_0 = \frac{\Gamma_2}{\Gamma}$, $\mathcal N_1 = \frac{\Gamma_2}{\Gamma} \frac{\Gamma_1}{\Gamma}$, and $\mathcal N_2 = \frac{\Gamma_2}{\Gamma} (\frac{\Gamma_1}{\Gamma})^2$. We conclude that $\mathcal S_N(\Delta_2) = \mathcal S_0(\Delta_2)$ for all values of $N$, and $\mathcal N_N =\frac{\Gamma_2}{\Gamma} (\frac{\Gamma_1}{\Gamma})^N$. Using Eq.\,\eqref{eq:spectrum}, the full spectrum results to be
\begin{equation}\label{spectrum}
	\mathcal S(\Delta_2) = \mathcal S_0(\Delta_2)\,. 
\end{equation}
The important conclusion is that the unknown number of previously scattered photons on the other transition is not observed as spectral broadening. Nevertheless, it is obvious that the incoherent summation of the various scattering processes weighted with their probabilities, Eq.\,\eqref{eq:spectrum}, will degrade the purity of the photonic state (see footnote~\cite{Note1}), i.\,e.\ the final photon will not be Fourier limited. This impurity is therefore solely due to the temporal broadening of the photonic wave packet through repeated decay back to the initial state and re-excitation before the final photon is emitted (see appendix~\ref{app_stretching}). This temporal broadening, and thereby the impurity, increases with the branching ratio~$\Gamma_1:\Gamma_2$.

\section{\label{Example}Quantum Beats}
\label{sec:beats}

\begin{figure}[htb]
	\centering
	\includegraphics{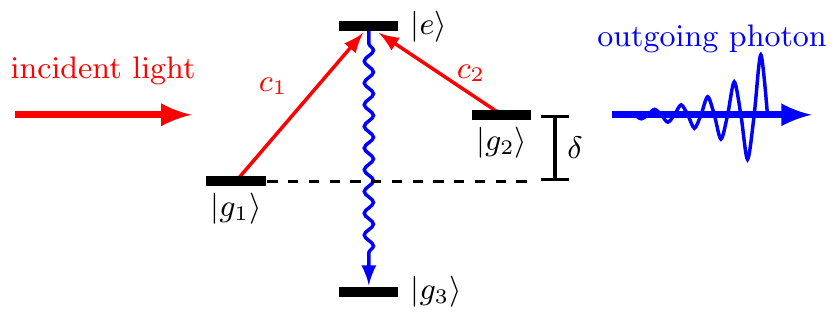}
	\caption{\label{fig:example} Atomic level configuration that leads to quantum beats. The atom is initially in a superposition of states~$\ket{g_1}$ and $\ket{g_2}$. A single photon is emitted on $\ket{e}\to\ket{g_3}$ after excitation by incident light.}
\end{figure}

The methods and results presented so far are extendable to more complex systems. As an example, we describe their application to an atomic level configuration that exhibits quantum beats in single-photon scattering~\cite{Schug2014}. Figure~\ref{fig:example} shows the level scheme with the relevant transitions. The atom is initially prepared in a superposition state of $\ket{g_1}$ and $\ket{g_2}$, and both transitions $\ket{g_1} \leftrightarrow \ket{e}$ and $\ket{g_2} \leftrightarrow \ket{e}$ are driven by the incident light. We are interested in the spectrum~$\mathcal S(\omega_3)$ of the photon emitted on the $\ket{e} \to \ket{g_3}$ transition. As before, we focus on the fully completed emission process, i.\,e.\ we set the interaction time $t \to \infty$.

To be specific, we consider the $\smash{^{40}\mathrm{Ca}^+}$ ion as in ref.\,\cite{Schug2014} whereby $\ket{g_1}$ and $\ket{g_2}$ are two Zeeman-split sub-levels of the $\mathrm D_{5/2}$ manifold, $\ket{e}$ is the $\mathrm P_{3/2}$ state, and $\ket{g_3}$ is the $\mathrm S_{1/2}$ ground state. Correspondingly, we use the branching fractions~$\Gamma_1/\Gamma = \Gamma_2/\Gamma = 0.03$ and $\Gamma_3/\Gamma = 0.94$ \footnote{In reality, the branching fractions will also contain the squares of the corresponding Clebsch--Gordan coefficients, but for conceptual clarity, this is omitted here.}. The frequency splitting between $\ket{g_1}$ and $\ket{g_2}$ is denoted as~$\delta$.

\subsection{Excitation by a single photon}

\begin{figure}[htb]
	\centering
	\includegraphics{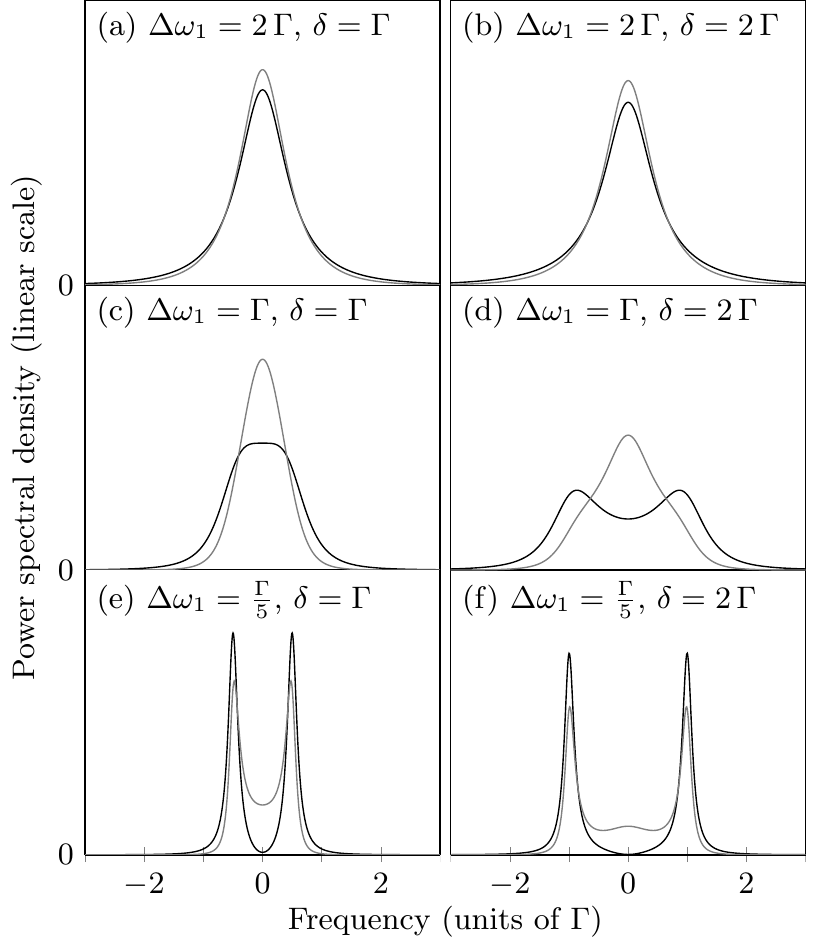}
	\caption{\label{fig:paketbeat}Area-normalized power spectral density~$\mathcal S(\Delta_3)$ for excitation of an initial superposition state by a single photon tuned to the center of the two transitions ($\Delta_1 = -\frac{\delta}{2}$). The incident wave packet is Lorentzian with width~$\Delta\omega_1$ as indicated; the splitting~$\delta$ is $\Gamma$ in (a), (c), and (e), and $2\,\Gamma$ in (b), (d), and (f). The gray levels indicate the phase between the two transitions: black for $c_1 = c_2$ and gray for $c_1 = -c_2$.}
\end{figure}

First, we assume excitation by a single photon with spectral amplitude~$\psi(\omega_1)$. The initial state is
\begin{equation*}
\ket{\Psi(0)} = \left( c_1 \ket{g_1} + c_2 \ket{g_2} \right) \otimes \int\limits_0^\infty \diff\omega_1 \psi(\omega) b_1^+(\omega_1)\ket{\text{vac}},
\end{equation*}
with $|c_1|^2 +|c_2|^2=1$. The final state after single-photon emission at frequency~$\omega_3$ is $\ket{f}=b_3^+(\omega_3)\ket{g_3; \text{vac}}$.
Since the absorption amplitudes of the two $\Lambda$-configurations add up coherently, the spectrum according to Eq.\,\eqref{eq:Somegat} is
\begin{equation}
\mathcal S(\Delta_3)= \frac{1}{\mathcal N} \left| c_1 U_{fi} \left( \Delta_1,\Delta_3 \right)+ c_2 U_{fi} \left( \Delta_1+\delta,\Delta_3 \right) \right|^2~. \label{beatspec}
\end{equation} 
The individual contributions~$U_{fi}$ are calculated as before using Eq.\,\eqref{eq:P} and \eqref{eq:ufi} for $t \to \infty$ and the spectral amplitude~$\psi(\omega_1)$. The emission spectrum for excitation by a Lorentzian photon 
is shown in fig.\,\ref{fig:paketbeat}. To highlight the case of equal transition strengths, we set $|c_1| = |c_2|$ and the detuning to $\Delta_1 = -\frac{\delta}{2}$, i.\,e.\ to the center between the two transitions. If the combined linewidth of the atomic response and input spectrum is smaller than the splitting of the two initial states, one observes two interfering spectral components, i.\,e.\ quantum beats, see fig.\,\ref{fig:paketbeat}(d-f). 

\subsection{Laser excitation}

\begin{figure}[htb]
	\centering
	\includegraphics{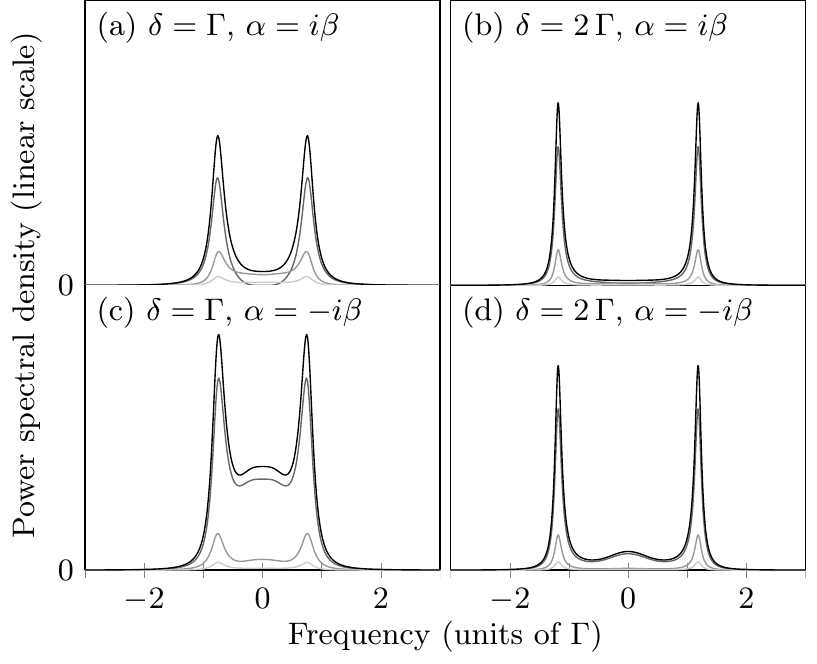} 
	\caption{\label{fig:laserbeat}Partial spectra~$\mathcal S_N(\Delta_3)$ for excitation by a laser tuned to the center of the two transitions ($\Delta_1 = -\frac{\delta}{2}$). The Rabi frequency is~$\Omega=\Gamma$. The splitting~$\delta$ is $\Gamma$ in (a) and (c), and $2\,\Gamma$ in (b) and (d). The gray levels indicate the number~$N$ of intermediate photons: dark, medium, and light gray for $N = 0$,~1,~2, respectively, and black for the sum.}
\end{figure}

For excitation by a laser with frequency~$\omega_1$, the fractions $\mathcal S_N(\Delta_3)$ of the spectral density corresponding to $N$~intermediately scattered photons (i.\,e.\ back to $\ket{g_{1,2}}$) are calculated as before (Sec.\,\ref{Laser}), but taking into account the coherent sum of Eq.\,(\ref{beatspec}). Notably, in this case the partial spectra are not identical in shape. Examples are plotted in fig.\,\ref{fig:laserbeat} for various splittings~$\delta$ and phases between $c_1$ and~$c_2$.

The presence of two peaks in the spectrum indicates quantum beats in the emitted photon with their difference frequency. In the case of laser excitation, the beat frequency contains a contribution from the AC Stark shift that becomes visible when the Rabi frequency approaches the atomic linewidth. The phase difference between the two transitions, $\arg c_1-\arg c_2$, also affects their interference, leading to pronounced maxima or minima between the two peaks. The corresponding changes in the temporal shape of the emitted photon have been reported in ref.\,\cite{Schug2014}.

\section{Conclusions}\label{sec:conclusion}

We have presented a detailed analysis of the spectra of single photons that are generated by a Raman-scattering process from a single quantum emitter, such as a trapped single ion or atom. Starting from basic principles we have calculated such spectra for various cases of excitation, in particular, by single incoming photons and by laser light. Our results back up previous experimental demonstrations where single photons of controlled temporal structure have been generated by laser excitation of a trapped ion~\cite{Keller2004, Kurz2013}, a trapped neutral atom~\cite{Darquie2005, Nisbet2013, Leong2016}, or where controlled single-photon Raman scattering has been utilized to implement a bi-directional atom--photon quantum interface~\cite{Kurz2016}. 

In general, our calculations show that the spectrum of an emitted photon is obtained as the product of the spectra of incoming light and atomic response. This includes the possible formation of dressed states in the atom if the excitation power approaches or exceeds saturation. In the temporal regime, the response of the atom and the incoming wave packet are convoluted to yield the outgoing wave packet. An important finding is that variation of the branching fractions of the excited atomic state, i.\,e.\ possible decay back to the initial state before the single Raman photon is released, does not affect the spectrum of the emitted photon. Only its temporal shape will be stretched, and hence the purity of the quantum state will degrade by a factor proportional to the average number of additional photons scattered on the excited transition. 

Beyond addressing a fundamental question in matter--light interaction, our analysis finds highly relevant applications in quantum communication technologies where single photons serve as carriers of quantum information. In this context their spectra determine, for example, the efficiency with which this information is transferred to atomic memories or the (in)distinguishability of two interfering photons. For the latter question, the purity of a photon's quantum state, i.\,e.\ how close its temporal and spectral representations are to the Fourier transforms of each other, is an important figure of merit. Our analysis shows how this purity depends on the atomic properties and the excitation parameters. 

Finally, we have extended our analysis to the case where the atom is initially in a superposition state for which quantum beats are observed and confirmed the findings of a recent experiment~\cite{Schug2014}.

\acknowledgements
We gratefully acknowledge financial support by the German Ministry of Education and Research (BMBF) through the Q.com-Q project (16KIS0127); we thank C.\ Kurz for helpful remarks.

P.\,M. and T.\,T. contributed equally to this work.

\appendix
\section{Laser excitation}\label{app_laser}

In the case of laser excitation, the matrix element of the time-evolution operator, Eq.\,\eqref{eq:Ufi}, has to be calculated individually for each case of $N$~additionally scattered photons. The matrix element of the resolvent, $G_{fi}(z)$, shows resonances at each involved state, thus $U_{f_Ni}(t)$ is a sum of just as many terms. Here, due to the interaction~$V$ (see Eq.\,\eqref{eq:V}), $\ket1$ and $\ket e$ are replaced by dressed states with the complex eigenfrequencies
\begin{equation*}\textstyle
	\omega_\pm= \frac{1}{2} \left( \Delta_1-i\frac{\Gamma}{2} \right) \pm \frac{1}{2} \sqrt{|\Omega|^2 +\left( \Delta_1+i\frac{\Gamma}{2} \right)^2}.
\end{equation*}

For $N=0$, there are three involved states with eigenfrequencies~$\omega_+$, $\omega_-$, and $\Delta_2$, leading to
\begin{multline}\label{eq:Uf0i}\textstyle
	U_{f_0 i}(t) = \frac{\Omega}{2} g_{\omega_2} e^{i \omega_1 t} \Bigg[ \frac{\displaystyle e^{-i\Delta_2 t}}{\left( \Delta_2 - \omega_+ \right)\left( \Delta_2-\omega_- \right)}\\
	\textstyle
	+\frac{\displaystyle e^{-i \omega_+ t}}{\left(\omega_--\omega_+ \right)  \left( \Delta_2 -\omega_+ \right)} +\frac{\displaystyle e^{-i\omega_- t}}{\left( \omega_+ - \omega_- \right)\left( \Delta_2 -\omega_-\right)} \Bigg].
\end{multline}
Since $\Im(\omega_\pm)<0$, only the first term in Eq.\,\eqref{eq:Uf0i} remains in the limit $t\to\infty$. It is the product of two Lorentzians centered at $\Re(\omega_\pm)$. By introducing the frequency,
\begin{equation*}
	\Delta_S = -\frac{\Delta_1}{2} + \frac{\text{sgn}(\Delta_1)}{2} \Re(\omega_+-\omega_-),
\end{equation*}
we get $\Re(\omega_\pm) = \frac{\Delta_1}{2} \pm \text{sgn}(\Delta_1) \left( \frac{\Delta_1}{2}+\Delta_S \right)$, such that one Lorentzian is centered at $\Delta_1+\Delta_S$ and the other one is centered at $-\Delta_S$. Their widths follow from $\Im(\omega_\pm)$ and are found to be $\kappa$ and $\Gamma-\kappa$, respectively. The resulting spectral density reads
\begin{multline}\label{spectrum0}
	\mathcal{S}_0(\Delta_2) = \frac{1}{\mathcal N_0} \left|U_{f_0 i}(\infty) \right|^2 \\
	= \frac{\frac{|\Omega|^2}{4} \frac{\Gamma}{2\pi} }{\left| \left( \Delta_2-\Delta_1-\Delta_S+i\frac{\kappa}{2} \right)\left( \Delta_2+\Delta_S+i\frac{\Gamma-\kappa}{2} \right) \right|^2}\,.
\end{multline}
The success probability~$\mathcal N_0 = \frac{\Gamma_2}{\Gamma}$ confirms that the probability of immediate decay of the excited state $\ket{e}$ to state $\ket{2}$ is just the corresponding branching fraction.

For $N=1$ there are five involved states with the eigenfrequencies~$\omega_+$, $\omega_-$, $\omega_++\Delta_1'$, $\omega_-+\Delta_1'$, and $\Delta_2+\Delta_1'$, (with $\Delta_1'=\omega_1'-\omega_1$) leading to
\begin{widetext}
\begin{multline}\label{eq:Uf1i}\textstyle
	U_{f_1i}(t) = \frac{\Omega^2}{4} g_{\omega_1'} g_{\omega_2} e^{i \omega_1 t}
	\left[ \frac{\displaystyle e^{-i\left( \Delta_2+\Delta_1' \right)t}}{\left( \Delta_2+\Delta_1' -\omega_+ \right)\left( \Delta_2+\Delta_1' -\omega_-\right) \left( \Delta_2-\omega_+\right)\left( \Delta_2 -\omega_-\right)}+
	\frac{\displaystyle e^{-i\omega_+ t}}{\left( \omega_+-\Delta_2-\Delta_1' \right) \left( \omega_+-\omega_-\right) \left(-\Delta_1' \right) \left( \omega_+-\omega_- -\Delta_1' \right)} \right.\\
	\textstyle
	+\left.\frac{\displaystyle e^{-i\omega_- t}}{\left( \omega_- -\Delta_2-\Delta_1' \right) \left( \omega_--\omega_+\right) \left( \omega_--\omega_+-\Delta_1' \right) \left( -\Delta_1' \right)}+
	\frac{\displaystyle e^{-i\left( \omega_++\Delta_1'  \right)t}}{\left( \omega_+-\Delta_2\right) \Delta_1' \left( \omega_+-\omega_-+\Delta_1' \right) \left( \omega_+-\omega_-\right)} +
	\frac{\displaystyle e^{-i\left( \omega_-+\Delta_1'\right)t}}{\left( \omega_- -\Delta_2 \right) \left( \omega_--\omega_+ +\Delta_1' \right) \Delta_1'  \left( \omega_--\omega_+\right) } \right].
\end{multline}
\end{widetext}
As before, only the first term in Eq.\,\eqref{eq:Uf1i} remains in the limit of $t\to\infty$, leading to the spectral density,
\begin{equation*}
	\mathcal{S}_1(\omega_2,t) = \frac{1}{\mathcal N_1} \int\limits_{-\infty}^\infty \diff{\omega_1'} \left|U_{f_1 i}(t) \right|^2,
\end{equation*}
where the integration over all possible values of $\omega_1'$ follows from the assumption that any information about the frequency of the additional photon is discarded. For $t\to\infty$, the integration leads to\begin{equation}\label{spectrum1}
	\mathcal{S}_1(\Delta_2) = \mathcal S_0(\Delta_2)\,.
\end{equation}
The spectra without and with an intermediate photon are the same because integration over all possible frequencies of the intermediate photon cancels out any correlation with the final one. The event $N=1$ has success probability
\begin{equation*}
	\mathcal N_1 = \frac{\Gamma_1\Gamma_2}{\Gamma^2}.
\end{equation*}

For $N=2$ there are seven involved states, so that the matrix element of the time-evolution operator, corresponding to Eq.\,\eqref{eq:Uf0i} and \eqref{eq:Uf1i}, is a sum of seven terms, each with six factors in the denominator,
\begin{widetext}
\begin{multline}\label{eq:Uf2i}\textstyle
U_{f_2i}(t) = \frac{\Omega^3}{8} g_{\omega_1'} g_{\omega_1''} g_{\omega_2} e^{i\omega_1 t} \left[\frac{\displaystyle e^{-i\left(\Delta_2+\Delta_1'+\Delta_1'' \right)t}}{\left(\Delta_2+\Delta_1'+\Delta_1''-\omega_+ \right)  \left(\Delta_2+\Delta_1'+\Delta_1''-\omega_- \right) \left( \Delta_2+\Delta_1''-\omega_+ \right) \left( \Delta_2+\Delta_1''-\omega_- \right) \left( \Delta_2-\omega_+ \right) \left( \Delta_2-\omega_-\right)} \right. \\
\textstyle
+ \frac{\displaystyle e^{-i\omega_+ t}}{\left( \omega_+-\Delta_2-\Delta_1'-\Delta_1'' \right) \left( \omega_+-\omega_- \right) \left(-\Delta_1' \right) \left(\omega_+-\omega_--\Delta_1' \right) \left( -\Delta_1'-\Delta_1'' \right) \left( \omega_+-\omega_--\Delta_1'-\Delta_1'' \right) } \\
\textstyle
+\frac{\displaystyle e^{-i\omega_- t}}{\left( \omega_- - \Delta_2-\Delta_1'-\Delta_1'' \right) \left( \omega_--\omega_+ \right) \left( \omega_--\omega_+-\Delta_1' \right) \left( -\Delta_1'\right) \left( \omega_--\omega_+-\Delta_1'-\Delta_1'' \right) \left( -\Delta_1'-\Delta_1'' \right)} \\
\textstyle
+ \frac{\displaystyle e^{-i\left(\omega_++\Delta_1' \right)t}}{\left( \omega_+-\Delta_2-\Delta_1'' \right) \left( \Delta_1' \right) \left( \omega_+-\omega_-+\Delta_1' \right) \left( \omega_+-\omega_- \right) \left( -\Delta_1'' \right) \left( \omega_+-\omega_--\Delta_1''\right)} 
\textstyle
+ \frac{\displaystyle e^{-i\left( \omega_-+\Delta_1' \right)t}}{\left( \omega_--\Delta_2-\Delta_1'' \right) \left(\omega_--\omega_++\Delta_1' \right) \left( \Delta_1' \right) \left( \omega_--\omega_+\right) \left(\omega_--\omega_+-\Delta_1'' \right) \left( -\Delta_1''\right)} \\
\textstyle
+\frac{\displaystyle e^{-i\left(\omega_++\Delta_1'+\Delta_1''\right)t}}{\left( \omega_+-\Delta_2\right)\left( \Delta_1'+\Delta_1'' \right) \left( \Delta_1''\right) \left( \omega_+-\omega_-+\Delta_1'+\Delta_1''\right) \left( \omega_+-\omega_-+\Delta_1''\right) \left( \omega_+-\omega_-\right)} 
\textstyle
+ \left.\frac{\displaystyle e^{-i\left( \omega_-+\Delta_1'+\Delta_1'' \right)t}}{\left( \omega_--\Delta_2 \right) \left( \omega_--\omega_++\Delta_1'+\Delta_1'' \right) \left(\Delta_1'+\Delta_1''\right) \left( \omega_--\omega_++\Delta_1'' \right) \left( \Delta_1''\right) \left( \omega_--\omega_+ \right)}   \right],
\end{multline}
\end{widetext}
where $\Delta_1''=\omega_1''-\omega_1$. Also here, only the first term in Eq.\,\eqref{eq:Uf2i} survives in the limit of $t\to\infty$, leading to the spectral density,
\begin{equation*}
	\mathcal{S}_2(\omega_2,t) = \frac{1}{\mathcal N_2} \int\limits_{-\infty}^\infty \diff{\omega_1'} \int\limits_{-\infty}^\infty \diff{\omega_1''} \left|U_{f_2 i}(t) \right|^2
\end{equation*}
which again simplifies to
\begin{equation}\label{spectrum2}
	\mathcal S_2(\Delta_2) = \mathcal S_0(\Delta_2),
\end{equation}
with
\begin{equation*}
	\mathcal N_2 = \frac{\Gamma_2}{\Gamma} \left(\frac{\Gamma_1}{\Gamma}\right)^2.
\end{equation*}

Generalization to all values of $N$ is obvious.

\section{Temporal stretching}\label{app_stretching}

To quantify the temporal distribution, or wave packet, of the finally emitted photon in the case of laser excitation, we consider that, after each spontaneously emitted photon on the transition $\ket{1}\leftrightarrow\ket{e}$, the emitter is projected back into the initial state, $\ket{1}$, and the excitation process starts again. For the first emitted photon (on either transition) after such a projection, the temporal shape of the wave packet is $p_1(t)$. As this distribution describes the uncertain moment of emission of the first photon, the wave packet of the second photon is broadened, i.\,e.\ its temporal shape, $p_2(t)$, is the convolution of the temporal shape of the first photon with itself. (Note that we trace out any possible interference between spontaneously emitted photons.) 


Consequently, the temporal shape of the $N$th photon is the convolution of the previous one with $p_1(t)$,
\begin{equation*}
	p_N(t) = (p_1*p_{N-1})(t).
\end{equation*}

In the case of a convolution of probability distributions one finds for the first moment,
\begin{equation*}
	\mean{t}_N = \int p_N(t)t\diff t = \mean{t}_{N-1} + \mean{t}_{1},
\end{equation*}
and the second central moment,
\begin{equation*}
	(\Delta t)_N^2 = \mean{(t-\mean{t}_N)^2}_N = (\Delta t)_{N-1}^2 + (\Delta t)_{1}^2.
\end{equation*}
Thus, the $N$th photon has the mean arrival time $\mean{t}_N = N\mean{t}_1$ and a temporal spread of $(\Delta t)_N = \sqrt{N}(\Delta t)_{1}$. 

%
%

From this we finally find the mean arrival time of the Raman photon by summing over all cases of $N$ additionally emitted photons weighted by their probabilities (cf.\ Eq.\,\eqref{eq:spectrum}),
\begin{equation*}
	\mean{t}_\text{Raman} = \sum_{N=0}^\infty \mathcal N_N\mean{t}_{N+1} = (\bar{N}+1)\mean{t}_1,
\end{equation*}
where $\bar{N}+1=\frac{\Gamma}{\Gamma_2}$ is the mean number of spontaneously emitted photons including the final one. By the corresponding calculation we find that the spread of the arrival time of the Raman photon is given by 
\begin{equation*}
	(\Delta t)_\text{Raman} = (\bar{N}+1)(\Delta t)_{1}.
\end{equation*}

\bibliography{pscatter}

\end{document}